\documentclass[aps,twocolumn,superscriptaddress]{revtex4-2}
\usepackage[colorlinks=true,bookmarks=true,citecolor=magenta,urlcolor=magenta,linkcolor=magenta,breaklinks]{hyperref} % command used
\usepackage{breakurl}
\usepackage{sidecap}
\usepackage{amssymb}
\usepackage{hhline}
\usepackage{urwchancal}
\usepackage{xcolor}
\sidecaptionvpos{figure}{t}
\usepackage{amsmath}
\usepackage{graphicx}
\usepackage{esint}
\usepackage{epstopdf}%This line makes .eps figures into .pdf - please comment out if not required.
\usepackage{rotating}
\graphicspath{{pict/}{./}}
\usepackage{bm}%
\newcounter{Fig}

\newcommand\mymapstol{\mathrel{\ooalign{$\leftarrow$\cr%
  \kern1.75ex\raise0.275ex\hbox{\scalebox{1}[0.4]{$\mid$}}\cr}}}

\newcommand\mymapstor{\mathrel{\ooalign{$\rightarrow$\cr%
  \kern-.15ex\raise.275ex\hbox{\scalebox{1}[0.4]{$\mid$}}\cr}}}
\begin{document}

%\titlefigure{abstract}

\title{Categorize coalescing quasi-normal modes through far-field scattering patterns}
\author{Jingwei Wang}
\affiliation{School of Optical and Electronic Information, Huazhong University of Science and Technology, Wuhan, Hubei 430074, P. R. China}
%\author{Pengxiang Wang}
%\affiliation{School of Optical and Electronic Information, Huazhong University of Science and Technology, Wuhan, Hubei 430074, P. R. China}
\author{Yuntian Chen}
\email{Email: yuntian@hust.edu.cn}
\affiliation{School of Optical and Electronic Information, Huazhong University of Science and Technology, Wuhan, Hubei 430074, P. R. China}
\affiliation{Wuhan National Laboratory for Optoelectronics, Huazhong University of Science and Technology, Wuhan, Hubei 430074, P. R. China}
\author{Wei Liu}
\email{Email: wei.liu.pku@gmail.com}
\affiliation{College for Advanced Interdisciplinary Studies, National University of Defense Technology, Changsha 410073, P. R. China.}
\affiliation{Nanhu Laser Laboratory and Hunan Provincial Key Laboratory of Novel Nano-Optoelectronic Information Materials and
Devices, National University of Defense Technology, Changsha 410073, P. R. China.}

\begin{abstract}
%We study chiroptical scattering properties of magneto-optical structures exhibiting rotational symmetry, with both rotation symmetry axis and externally applied magnetic field ($\mathbf{B}_{\rm{ext}}$)  being parallel to the incident direction.  For more than $3$-fold rotational symmetries, irrelevant to $\mathbf{B}_{\rm{ext}}$, scattering and 
%extinction cross sections (${C}_{\rm{sca}}$ and ${C}_{\rm{ext}}$) are invariant as along as the incident polarizations locate on the same latitude line of the Poincaré sphere (polarization ellipses of the same ellipticity while different orientations).  While for general incident polarizations, ${C}_{\rm{ext}}$ and ${C}_{\rm{sca}}$ are ellipticity-weighted arithmetic average of those for circularly-polarized incident waves. 
Resonances in the form of quasi-normal modes (QNMs) for open scattering systems can be generally identified in the far field through peaks of scattering spectra (\textit{e.g.} cross sections of scattering, extinction and absorption). Nevertheless, when the resonant frequencies of different QNMs are spectrally overlapped or sufficiently close,  the scattering peaks merge and it then becomes extremely challenging to reveal the mode constituents underlying the scattering spectra solely in the far field. Here in this work, we study open systems scattering electromagnetic plane waves, and reveal that spectrally close or even overlapped QNMs can be selectively excited through tuning incident directions and polarizations. Such a far-field technique can be further applied to categorize the nature of degenerate QNMs sharing identical complex eigenfrequencies (coalescent eigenvalues): at effectively Hermitian degeneracies (conical or Dirac points),  the eigenvectors are not coalescent and thus the QNMs can still be selectively excited, producing distinct scattering patterns; while for non-Hermitian degeneracies (exceptional points), eigenvectors also coalesce and thus selective QNM excitation does not exist, leading to invariant scattering patterns. Our technique sheds new light on the borderlands of Mie theory, QNMs, non-Hermitian photonics and singular optics, which can empower new explorations and applications and cross-fertilize all those disciplines. 
\end{abstract}

\maketitle

\section{Introduction}
\label{section1}

Open scattering systems generally support quasi-normal modes (QNMs) that are characterized by complex eigenfrequencies and eigenvectors (vectorial electromagnetic field distributions)~\cite{LALANNE__LaserPhotonicsRev._Light}.  The scattering properties (such as scattering and extinction cross sections~\cite{Bohren1983_book}) of those systems originate from the excitations and interferences of the QNMs excited, and the excitation coefficients are dependent on the morphologies of incident fields (temporal and spatial phase and polarization distributions)~\cite{LALANNE__LaserPhotonicsRev._Light,Bohren1983_book,POWELL_Phys.Rev.Applied_interference_2017,WANG_2024_Phys.Rev.Lett._Geometric}. Generally speaking, QNMs can manifest themselves through peaks in the scattering spectra, and central positions and linewidths of those peaks approximately correspond to the real parts of eigenfrequencies and  Q-factors of the QNMs excited~\cite{LALANNE__LaserPhotonicsRev._Light,POWELL_Phys.Rev.Applied_interference_2017}. Nevertheless, such a correspondence would break down when two or more QNMs are spectrally close or even overlapped, rendering several peaks merged together, \textit{e.g.} inducing asymmetric line shapes of Fano resonances~\cite{LIMONOV_NatPhoton_fano_2017}).  To reveal
the details of QNM excitations and modal contributions
underlying those compound merged peaks, 
near-field or far-field techniques involving volume or contour integrations can be employed~\cite{LALANNE__LaserPhotonicsRev._Light,BINKOWSKI_2020_Phys.Rev.B_Quasinormal}. The problem is that from those integrations intuitive physical insights and general principles are hard to extract to directly and precisely tell how to tailor the incident sources for flexible selective mode excitations,    which are of great significance for manipulations of scattering properties.

Different QNMs are generally featured by distinct near-field distributions, and different far-field radiation patterns in terms of both angular intensities and polarization distributions on the momentum sphere~\cite{LALANNE__LaserPhotonicsRev._Light}. For a reciprocal scatterer, if there exists a radiation direction along which any pair of QNMs have different radiation polarizations, it is recently revealed that the two QNMs can be selectively excited for plane waves incident opposite to such radiation direction with proper polarizations~\cite{WANG_2024_Phys.Rev.Lett._Geometric,CHEN_Phys.Rev.Lett._Extremize,ZHANG_2025_OpticsCommunications_Selective}. When a specific QNM is exclusively excited, the far-field scattering pattern would be identical to this mode radiation pattern, which serves as an indicator for selective QNM excitations~\cite{ZHANG_2025_OpticsCommunications_Selective}. 

In this study, we  employ this approach based on far-field scattering patterns to categorize QNMs coalescences into: (i) QNMs with identical real (distinct imaginary) parts of eigenfrequencies; (ii) Effectively Hermitian degenerate (with identical complex eigenfrequencies) QNMs with non-coalescent eigenvectors; (iii) Effectively non-Hermitian degenerate  QNMs  with coalescent eigenvectors. For categories of (i) \& (ii), though spectrally overlapped, the distinct QNMs can be selectively excited with designed incident directions and polarizations,  leading to different scattering patterns dictated by radiation patterns of the corresponding QNMs. While for the last category of (iii),  effectively only one QNM exists, and thus there is no selective mode excitation and thus the scattering pattern would be invariant for varying incident polarizations.  Our far-field approach to distinguish among coalescent QNMs has merged sweeping concepts of Mie scattering, QNMs, singularities, and (non-)Hermitian degeneracies, which can play significant roles for not only electromagnetic wave manipulations, but also waves of other forms where those concepts are generic and ubiquitous.

\section{Theoretical model for selective QNM excitation.}
\label{section2}
In this work, we study reciprocal open scattering systems that support a discrete set of QNMs (indexed by integer $n$), for which the vectorial eigenfield and complex eigenfrequency are respectively $\tilde{\mathbf{E}}_{n}(\mathbf{r})$ and $\tilde{\omega}_{n}$. We have further confined the incident waves to plane waves with electric vector field ${{\mathbf{E}}}_{\rm{inc}}(\mathbf{\hat{r}}_{\rm{inc}})$ and angular frequency $\omega_{\rm{inc}}$, where  $\mathbf{\hat{r}}_{\rm{inc}}$ is the unit incident direction vector.  Along an arbitrary radiation direction (unit direction vector $\mathbf{\hat{r}}_{\rm{rad}}$), in the far field the QNM radiations are transverse $\tilde{\mathbf{E}}_{n}(\mathbf{\hat{r}}_{\rm{rad}})\cdot\mathbf{\hat{r}}_{\rm{rad}}=0$. The scattered far field can be expanded as coherent superpositions of the radiations of all QNMs excited~\cite{LALANNE__LaserPhotonicsRev._Light}:
%%%====================================
\begin{equation}
\label{expansion}
\mathbf{E}_{\rm{sca}}(\mathbf{\hat{r}}_{\rm{sca}})=\sum \alpha_{n} {\tilde{\mathbf{E}}}_{n}(\mathbf{\hat{r}}_{\rm{sca}}),  
\end{equation}
%%%===================================
where $\alpha_{n}$ is the complex excitation (expansion) coefficient for the $n^\mathrm{th}$ QNM; that is, $|\alpha_{n}|^2$ is the excitation efficiency for the $n^\mathrm{th}$ QNM;   $\mathbf{\hat{r}}_{\rm{sca}}$ is the unit scattering direction vector. 

For reciprocal open scattering bodies excited by incident plane waves, the QNM excitation coefficient is solely related to the QNM radiation opposite to the incident direction~\cite{WANG_2024_Phys.Rev.Lett._Geometric,CHEN_Phys.Rev.Lett._Extremize,ZHANG_2025_OpticsCommunications_Selective}:

%%%====================================
\begin{equation}
\label{expansion-coefficient}
\alpha_n\propto\mathrm{\mathbf{E}}_{\rm{inc}}(\mathbf{\hat{r}}_{\rm{inc}})\cdot\mathrm{\tilde{\mathbf{E}}}^{\ast}_{n}(\mathbf{\hat{r}}_{\rm{rad}}=-\mathbf{\hat{r}}_{\rm{inc}}),
\end{equation}
%%%===================================
where  $\ast$  denotes complex conjugate. To fully suppress a specific QNM $\mathbf{A}$, it is only required that the incident polarization is orthogonal to this mode radiation polarization along $-\mathbf{\hat{r}}_{\rm{inc}}$:
%%%====================================
\begin{equation}
\label{expansion-coefficient-suppression}
\alpha_{\textbf{A}}=\mathrm{\mathbf{E}}_{\rm{inc}}(\mathbf{\hat{r}}_{\rm{inc}})\cdot\mathrm{\tilde{\mathbf{E}}}^{\ast}_{\mathbf{A}}(\mathbf{\hat{r}}_{\rm{rad}}=-\mathbf{\hat{r}}_{\rm{inc}})=0.
\end{equation}
%%%===================================
For an open system supporting two dominant QNMs  $\mathbf{A}$ and $\mathbf{B}$, as long as their radiation polarizations are distinct opposite to the incident direction ($\tilde{\mathbf{S}}$ is the Stokes vector that characterizes the polarization of $\tilde{\mathbf{E}}$~\cite{YARIV_2006__Photonics}):
\begin{equation}
\label{expansion-coefficient-polarization}
\mathrm{\tilde{\mathbf{S}}}_{\mathbf{A}}(\mathbf{\hat{r}}_{\rm{rad}}=-\mathbf{\hat{r}}_{\rm{inc}})\neq\mathrm{\tilde{\mathbf{S}}}_{\mathbf{B}}(\mathbf{\hat{r}}_{\rm{rad}}=-\mathbf{\hat{r}}_{\rm{inc}}),
\end{equation}
fully suppression of QNM $\mathbf{A}$ means selective excitation of QNM $\mathbf{B}$ (otherwise if $\mathrm{\tilde{\mathbf{S}}}_{\mathbf{A}}=\mathrm{\tilde{\mathbf{S}}}_{\mathbf{B}}$ both modes would be simultaneously excited or suppressed). Then according to  Eqs.~(\ref{expansion-coefficient}) and (\ref{expansion-coefficient-suppression}), the scattering pattern would be the same as the radiation pattern of the selectively excited mode. We emphasize that in both our approach and those in Refs.~\cite{LALANNE__LaserPhotonicsRev._Light,BINKOWSKI_2020_Phys.Rev.B_Quasinormal}, numerical calculations to obtain the properties of QNMs are required. The simplicity of our model [Eq.~(\ref{expansion-coefficient})] resides in that neither further simulations with the incident sources (as in Ref. \cite{BINKOWSKI_2020_Phys.Rev.B_Quasinormal}) nor evaluations of integrations (as in Ref.~\cite{LALANNE__LaserPhotonicsRev._Light}) are needed to calculate the mode excitation coefficients. It is exactly this simplicity that renders a direct recipe according to which we can identify the proper incident directions and polarizations to manipulate the QNM excitations.  Otherwise, using the techniques in Refs.~\cite{LALANNE__LaserPhotonicsRev._Light,BINKOWSKI_2020_Phys.Rev.B_Quasinormal}, further extensive calculations in four-dimensional parameter space
(two dimensions of the incident momentum sphere plus two dimensions of the
polarization Poincaré sphere) are required, besides the information of the QNM.

Up to now, it has become clear that for any pair of QNMs, they can be selectively excited (suppressed) as long as their radiation polarizations are distinct along the opposite incident directions. This principle of selectivity has nothing to do with QNM eigenfrequencies. That is, selective excitation is accessible if there exist one single direction along which the QNM radiations are of distinct polarizations. The exceptional scenario that escapes this net of selection is the non-Hermitian degenerate points (exceptional points), at which all eigenvectors are coalescent~\cite{BERRY_2004_CzechoslovakJournalofPhysics_Physicsa,MOISEYEV_2011__NonHermitian,WANG_2023_Adv.Opt.Photon._NonHermitian}, meaning that the QNM radiation polarizations are identical throughout the whole momentum sphere. This is exactly the key point we exploit to distinguish between effectively Hermitian and  non-Hermitian degeneracies of QNMs supported by open reciprocal scatterers.

\section{Mode excitation for coalescing QNMs}
\begin{figure}[tp]
\centerline{\includegraphics[width=8.5cm]{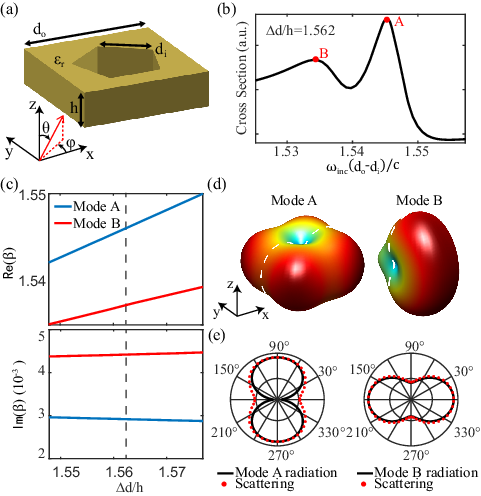}} \caption {\small (a) A dielectric ($\varepsilon_r=40$) square block (side length $d_o$ and height $h$) with a  centered regular-pentagon puncture (side length $d_i$). The Cartesian coordinate system is also parameterized by azimuthal angle $\varphi$ and polar angle $\theta$. (b) Scattering cross section spectrum for the left-handed circularly polarized plane wave incident along the -$\mathbf{y}$ direction, with ${\Delta d}/{h}=1.562$. (c) Normalized complex eigenfrequencies $\beta=\tilde{\omega}({d_{\text {o }}-d_{\text {i}}})/{c}$ (both real and imaginary  parts) versus geometric parameters  for the two supported QNMs \textbf{A} and \textbf{B} identified through the peaks in (b). (d) Radiation patterns (in terms of angular scattering intensity) for the two QNMs \textbf{A} and \textbf{B} with ${\Delta d}/{h}=1.562$. (e)  Scattering and mode radiation patterns on the \textbf{x}-\textbf{z} plane for waves incident along -\textbf{y} direction.   In (e), both scenarios with different incident polarizations are shown (incident polarizations are orthogonal respectively to that of modes \textbf{B} and \textbf{A} radiation polarizations along the  +$\mathbf{y}$ direction): mode \textbf{A} selectively excited with $\beta_{\rm{inc}}=\omega_{\mathrm{inc}}\left({d}_{{o}}-{d}_{{i}}\right) / \mathrm{c}=\textbf{Re}(\beta_A)$ (left); mode \textbf{B} selectively excited with $\beta_{\rm{inc}}=\textbf{Re}(\beta_B)$ (right). With ${\Delta d}/{h}=1.562$: $\beta_A=1.5458+2.9166 * 10^{-3}\mathrm{i}$; $\beta_B= 1.5370+4.4152 * 10^{-3} \mathrm{i}$; along the  +$\mathbf{y}$ direction the radiations of QNMs \textbf{A} and \textbf{B}  are linearly polarized along $\mathbf{z}$ and $\mathbf{x}$ axes, respectively. In this figure we have fixed $d_i/d_o=0.306$.}\label{fig1}
\end{figure}

To verify the principle of selectivity we have elaborated on in the last section, we then turn to specific open scattering bodies for numerical demonstrations, and all the following numerical results  are obtained using COMSOL Multiphysics. A scatterer and the coordinate system (with polar angle $\theta$ and azimuthal angle $\varphi$) are shown schematically in Fig.~\ref{fig1}(a): a square block (side length $d_o$ and height $h$) with a  centred regular-pentagon puncture (side length $d_i$); the scatterer is homogenous with isotropic relative permittivity $\varepsilon_r=40$ and fixed $d_i/d_o=0.306$. The scattering cross section spectrum for this scatterer [the incident direction vector $\mathbf{\hat{r}}_{\rm{inc}}$=($\theta_i$,  $\varphi_i$)=($\pi/2$,  $-\pi/2$); that is, the plane wave is incident along -\textbf{y} direction and  it is left-handed circularly polarized] is shown in Fig.~\ref{fig1}(b), where two scattering peaks are present.  Associated with each peak there is an individual QNM.  This is confirmed by Fig.~\ref{fig1}(c), where we show the dependence of complex QNM eigenfrequency $\tilde{\omega}$ [both real ($\mathbf{Re}$) and imaginary ($\mathbf{Im}$) parts are normalized: $\beta=\tilde{\omega}({d_{\text {o }}-d_{\text {i}}})/{c}$; $c$ is the speed of light on structural parameter ${\Delta d}/{h}=({d_{\text {o }}-d_{\text {i}}})/{h}$. It is clear that for the chosen parameter in Fig.~\ref{fig1}(b) (${\Delta d}/{h}=1.562$), there are two spectrally separated QNMs with both real and imaginary parts of eigenfrequencies being distinct [see  Fig.~\ref{fig1}(c)]. 

\begin{figure}[tp]
\centerline{\includegraphics[width=8.5cm]{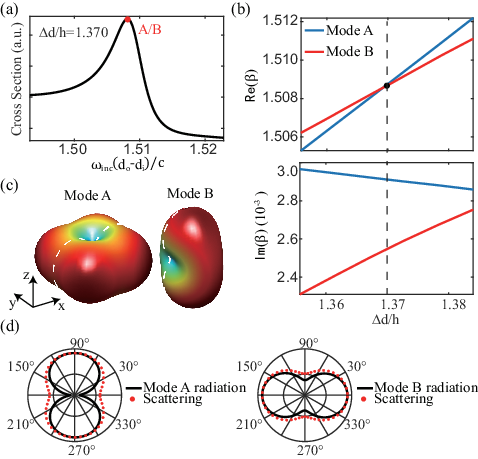}} \caption{\small (a) Scattering cross section spectrum for the left-handed circularly polarized plan wave incident along the -$\mathbf{y}$ direction, with ${\Delta d}/{h}=1.37$. (b) Normalized complex eigenfrequencies (both real and imaginary  parts) versus geometric parameters  for the two supported QNMs \textbf{A} and \textbf{B} underlying the peak in (a). (c) Radiation patterns (in terms of angular scattering intensity) for the two QNMs \textbf{A} and \textbf{B} with ${\Delta d}/{h}=1.37$. (d)  Scattering and mode radiation patterns on the \textbf{x}-\textbf{z} plane for waves incident along -\textbf{y} direction.   In (d), both scenarios with different incident polarizations are shown (incident polarizations are orthogonal respectively to that of modes \textbf{B} and \textbf{A} radiation polarizations along the  +$\mathbf{y}$ direction) with $\beta_{\rm{inc}}=\textbf{Re}(\beta_A)=\textbf{Re}(\beta_B)$:  mode \textbf{A} selectively excited (left) and mode \textbf{B} selectively excited (right). For ${\Delta d}/{h}=1.37$: $\beta_A=1.5088+2.9074 * 10^{-3} \mathrm{i}$; $\beta_B= 1.5088+2.5706 * 10^{-3} \mathrm{i}$; along the  +$\mathbf{y}$ direction the radiations of QNMs \textbf{A} and \textbf{B}  are linearly polarized along $\mathbf{z}$ and $\mathbf{x}$ axes, respectively.  In this figure we have fixed $d_i/d_o=0.294$.}
\label{fig2}
\end{figure}

\begin{figure}[tp]
\centerline{\includegraphics[width=8.5cm]{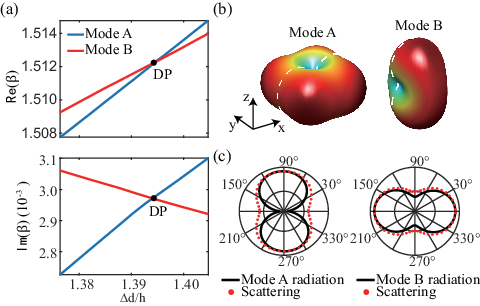}} \caption{\small (a) Normalized complex eigenfrequencies (both real and imaginary parts) versus geometric parameters  for  two supported QNMs, which become degenerate at ${\Delta d}/{h}=1.395$ (location of the DP). (b) Radiation patterns (in terms of angular scattering intensity) for the two QNMs \textbf{A} and \textbf{B} with ${\Delta d}/{h}=1.395$. (c)  Scattering and mode radiation patterns on the \textbf{x}-\textbf{z} plane for waves incident along -\textbf{y} direction.   In (c), both scenarios with different incident polarizations are shown (incident polarizations are orthogonal respectively to that of modes \textbf{B} and \textbf{A} radiation polarizations along the  +$\mathbf{y}$ direction) with $\beta_{\rm{inc}}=\textbf{Re}(\beta_A)=\textbf{Re}(\beta_B)$:  mode \textbf{A} selectively excited (left) and mode \textbf{B} selectively excited (right). For ${\Delta d}/{h}=1.395$: $\beta_A=\beta_B= 1.5122+2.9794 * 10^{-3} \mathrm{i}$; along the  +$\mathbf{y}$ direction the radiations of QNMs \textbf{A} and \textbf{B}  are linearly polarized along $\mathbf{z}$ and $\mathbf{x}$ axes, respectively.  In this figure we have fixed $d_i/d_o=0.296$.
}
\label{fig3}
\end{figure}

The far-field radiation patterns for QNMs \textbf{A} and \textbf{B} are shown in  Fig.~\ref{fig1}(d), and the mode radiation polarizations along +\textbf{y} direction (opposite to the incident direction) are linear polarizations along \textbf{z} and \textbf{x} axes, respectively.  As a result, according to Eqs.~(\ref{expansion-coefficient}) and (\ref{expansion-coefficient-suppression}), for a plane wave incident along  -\textbf{y} direction and linearly polarized along \textbf{z} (\textbf{x}) axes with $\alpha_\textbf{B}=\mathrm{\mathbf{E}}_{\rm{inc}}\cdot\mathrm{\tilde{\mathbf{E}}}^{\ast}_{\mathbf{B}}=0$ ($\alpha_\textbf{A}=\mathrm{\mathbf{E}}_{\rm{inc}}\cdot\mathrm{\tilde{\mathbf{E}}}^{\ast}_{\mathbf{A}}=0$ ), QNM \textbf{A} (\textbf{B}) would be selectively excited, with QNM \textbf{B} (\textbf{A}) fully suppressed. This is further verified in Fig.~\ref{fig1}(e), where we show the scattering and mode radiation patterns on the \textbf{x}-\textbf{z} plane parameterized by $\theta$. It is clear that the scattering patterns are almost identical to the radiation patterns of QNMs \textbf{A} and \textbf{B}, respectively,  confirming selective mode excitations. The discrepancies between the scattering and radiation patterns are induced by marginal contributions of other QNMs that are spectrally outside the spectral regime of interest, as is also the case for  
Figs.~\ref{fig2}-\ref{fig4}.

%%%%%%%%%%%%%%%%%%%%%%%%%%%%%%%
 %($\lambda_{A-G}/R=6.16,~9.54,~4.58,~3.54,~9.54,~3.54$, and $6.07$)
%%%%%%%%%%%%%%%%%%%%%%%%%%%%%%%%%%%

%====================================
%=====================================

\subsection{Two QNMs with identical $\mathbf{Re}(\tilde{\omega})$}
We further tune the geometric parameter ($d_i/d_o=0.294$) of the structure shown in Fig.~\ref{fig1}(a) to obtain two coalescing QNMs with the same real part of eigenfrequency. In the scattering spectrum shown in Fig.~\ref{fig2}(a) [incident polarization and direction are the same as those in Fig.~\ref{fig1}(b); ${\Delta d}/{h}=1.37$] there is only one peak, underlying which there are actually two rather than one QNMs. The complex eigenfrequency dependence on geometric parameter is summarized in Fig.~\ref{fig2}(b), confirming that  $\mathbf{Re}(\tilde{\omega})$ for the two QNMs is identical with distinct  $\mathbf{Im}(\tilde{\omega})$. The radiation patterns of the two QNMs are summarized in Fig.~\ref{fig2}(c) [similar to the patterns shown in Fig.~\ref{fig1}(d)]. Since the mode radiation polarizations along +\textbf{y} directions are also respectively linearly polarized along \textbf{z} and \textbf{x} axes, these two QNMs with identical $\mathbf{Re}(\tilde{\omega})$ can be selectively excited (with incident polarization orthogonal to the radiation polarization of the suppressed mode), as has been confirmed by the radiation and scattering patterns shown in  Fig.~\ref{fig2}(d).

%====================================
%=====================================

\subsection{Two QNMs with identical $\tilde{\omega}$: effective Hermitian degeneracy}
Open scattering systems can be tuned to support QNMs with identical $\tilde{\omega}$ in terms of both real and imaginary parts~\cite{DMITRIEV_2023_Phys.Rev.A_Retardationinduced,CANOSVALERO_2024_Phys.Rev.Research_Bianisotropic,SOLODOVCHENKO_2024_Phys.Rev.B_Quadruplets,ZHANG_2024__NonHermitian}. Though scattering systems are essentially non-Hermitian, the mode degeneracy can be effectively Hermitian (conical or Dirac point, DP): it means  that though the eigenvalues are coalescent, the eigenvectors are orthogonal rather than coalescent. Such a scenario is shown in Fig.~\ref{fig3}(a), employing the scatterer shown in Fig.~\ref{fig1}(a) with a different geometric parameter $d_i/d_o=0.296$. The radiation patterns of the two effectively Hermitian degenerate modes are shown in Fig.~\ref{fig3}(b), and their respective selective excitations have been confirmed by Fig.~\ref{fig3}(c).

%====================================
\begin{figure}[tp]
\centerline{\includegraphics[width=8.5cm]{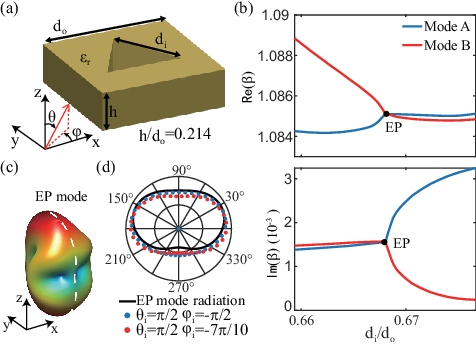}} \caption{\small (a) A dielectric ($\varepsilon_r=40$) square block (side length $d_o$ and height $h$) with a  centred regular-triangle puncture (side length $d_i$). (b) Normalized complex eigenfrequencies (both real and imaginary parts) versus geometric parameters  for  two supported QNMs, which become degenerate at ${d_i}/{d_o}=0.668$ (location of the EP). (c) Radiation patterns (in terms of angular scattering intensity) for the non-Hermitian degenerate QNMs with ${d_i}/{d_o}=0.668$. (d)  Scattering and mode radiation patterns on the \textbf{y}-\textbf{z} plane for linearly-polarized (along \textbf{z} axis) waves incident along two directions specified, with $\beta_{\rm{inc}}=\textbf{Re}(\beta_A)=\textbf{Re}(\beta_B)$. For ${d_i}/{d_o}=0.668$: $\beta_A=\beta_B= 1.0847+1.5640 * 10^{-3} \mathrm{i}$.  In this figure we have fixed $h/d_o=0.214$.}
\label{fig4}
\end{figure}
%=====================================

\subsection{Two QNMs with identical $\tilde{\omega}$: non-Hermitian degeneracy}
To obtain and demonstrate the scenario of non-Hermitian degeneracy (exceptional point, EP), we employ now a slightly different structure schematically shown in  Fig.~\ref{fig4}(a): a dielectric square block with a regular-triangle puncture and the same relative permittivity $\varepsilon_r=40$. The degeneracy in terms of identical $\tilde{\omega}$ is showcased in Fig.~\ref{fig4}(b). In sharp contrast to the DP (see Fig.~\ref{fig3}) at which the two QNMs are still different in both near and far fields, at EP there is effectively only one QNM [eigenvectors are also coalescent; the two QNMs merge into one QNM, of which the radiation pattern is shown in Fig.~\ref{fig4}(c)]. In other words, for arbitrary incident directions and polarizations, only one QNM is effectively excited and thus the scattering pattern becomes invariant. Such invariance is confirmed in Fig.~\ref{fig4}(d), for two different incident directions.

\section{Conclusion and Perspective}

To conclude, we develop a far-field technique to distinguish among different types of coalescing QNMs, covering QNMs with eigenfrequencies of identical real part, and degenerate QNMs of both effective Hermitian and non-Hermitian natures. For spectrally close or even overlapped QNMs, as long as the eigenvectors are not coalescent, QNMs can be selectively excited, inducing different far-field scattering patterns. While for non-Hermitian degenerate QNMs, since effectively there is only one QNM and its sole excitation, the scattering pattern is dictated by its radiation pattern, which is invariant for different incident scenarios. Our technique has effectively merged the pervading concepts of Mie scattering, singularity, QNMs, Hermitian and non-Hermitian degeneracies, which will stimulate novel interdisciplinary investigations in the borderlands of the vibrant fields of topological, singular and non-Hermitian photonics~\cite{WANG_2023_Adv.Opt.Photon._NonHermitian,OZAWA_2018_ArXiv180204173,NYE_natural_1999,LIU_ArXiv201204919Phys._Topological}. 

Here in this paper, we have discussed only a pair of coalescing QNMs. For more than two QNMs and their higher-order degeneracies, exclusively selective excitation of one specific mode and suppression of all other modes requires more than one plane waves incident along different directions with different polarizations. For other incident sources beyond plane waves, our technique is not directly applicable (if the incident source can be expanded into a series of plane waves, the functionalities of linear systems can be still obtained through response integrations based on our technique), and more comprehensive and generic near- or far-field techniques can be employed to decide the modal contributions~\cite{LALANNE__LaserPhotonicsRev._Light,BINKOWSKI_2020_Phys.Rev.B_Quasinormal}. Moreover, our formalisms presented in Section~\ref{section2} are valid only for reciprocal scattering bodies, and extra extensions are required to provide a more comprehensive framework for nonreciprocal systems with magnetism, nonlinearity or temporal modulations~\cite{FAN_science_comment_2012,CALOZ_2018_Phys.Rev.Applied_Electromagnetic}. Those extensions and generalizations can not only significantly broaden the horizons of aforementioned disciplines in photonics, but also bring new opportunities for waves of other forms, where the concepts of scattering, QNMs and singularity are generic ubiquitous.   \\

\section*{acknowledgement}
This work is supported by National
Natural Science Foundation of China (Grants
No. 11874426 and No. 61405067), Outstanding Young
Researcher Scheme of Hunan Province (2024JJ2056), and
several Researcher Schemes of National University of
Defense Technology. W. L. acknowledges fruitful discussions with A. A. Bogdanov and Y. S. Kivshar.

%Q. Yang and W. Chen contributed equally to this work.

%load xkeyval.sty

%\bibliographystyle{osajnl}
%\bibliography{References_scattering7}


\end{document}